# How Users Experience Closed Captions on Live Television: Quality Metrics Remain a Challenge

MARIANA ARROYO CHAVEZ*

Gallaudet University, mariana.arroyo.chavez@gallaudet.edu

MOLLY FEANNY

Gallaudet University, molly.feanny@gallaudet.edu

MATTHEW SEITA

Gallaudet University, matthew.seita@gallaudet.edu

BERNARD THOMPSON

Gallaudet University, bernard.thompson@gallaudet.edu

KEITH DELK

Gallaudet University, keith.delk@gallaudet.edu

SKYLER OFFICER

Gallaudet University, skyqinoff@gmail.com

ABRAHAM GLASSER

Gallaudet University, abraham.glasser@gallaudet.edu

RAJA KUSHALNAGAR

Gallaudet University, raja.kushalanagar@gallaudet.edu

CHRISTIAN VOGLER

Gallaudet University, christian.vogler@gallaudet.edu

This paper presents a mixed methods study on how deaf, hard of hearing and hearing viewers perceive live TV caption quality with captioned video stimuli designed to mirror TV captioning experiences. To assess caption quality, we used four commonly-used quality metrics focusing on accuracy: word error rate, weighted word error rate, automated caption evaluation (ACE), and its successor ACE2.

---

* The first two authors contributed equally to this paper.

We calculated the correlation between the four quality metrics and viewer ratings for subjective quality and found that the correlation was weak, revealing that other factors besides accuracy affect user ratings. Additionally, even high-quality captions are perceived to have problems, despite controlling for confounding factors. Qualitative analysis of viewer comments revealed three major factors affecting their experience: Errors within captions, difficulty in following captions, and caption appearance. The findings raise questions as to how objective caption quality metrics can be reconciled with the user experience across a diverse spectrum of viewers.

CCS CONCEPTS • **Human-centered computing~Accessibility~Empirical studies in accessibility**

**Additional Keywords and Phrases:** closed captioning, quality metrics, caption usability, subtitles

## 1 INTRODUCTION

Closed captions, or subtitles, provide visual access to the spoken content in videos, especially to deaf and hard of hearing (DHH) people. Access to video has a direct impact on societal participation, including entertainment, keeping up with news, political processes, contact with institutions and elected representatives, education, and acquiring new skills. Captioned videos not only aid people with varying levels of hearing loss, but also anyone with temporary or situational hearing loss; for example, patrons in noisy places, or in places with multiple televisions, such as bars or airports. The public often falsely assumes that captioned video programs provide full access to the underlying speech information [13]. This is not the case: captions are neither speech nor text, but rather come with their own considerations and editorial trade-offs [39][41].

In the United States, closed captioning on television broadcasts is regulated by the Federal Communications Commission (FCC). The FCC rules stipulate that nearly all TV programs in the United States must be captioned. However, while some countries, such as Canada and the United Kingdom, have adopted caption quality metrics, the FCC has not done so to date. Rather, it has defined a set of best industry practices, focused on four areas of quality – placement, synchronicity, accuracy, and completeness [15]. Overall, some countries have already adopted metrics while others haven't, in either case it is unclear if the metrics actually align with viewer expectations.

There is no consensus in the United States as to how to quantitatively measure captions in alignment with these four quality criteria. From a user perspective, caption quality generally cannot be evaluated in terms of simple metrics like accuracy or precision because there is no objective notion of ground truth by a caption reader, for they do not know what the speaker intended to say [13][40]. Furthermore, even hearing transcribers cannot say with 100% confidence what the speaker intended to say, and two separate individuals may transcribe the same audio source in different ways [41]. Therefore, the caption reader must use contextual understanding to quickly detect and repair errors mentally, regardless of whether it was misspoken by the speaker, misheard/mis-transcribed or misunderstood by the reader. Moreover, there is deep disagreement about what aspects of caption quality matter the most. For instance, there are trade-offs among word accuracy, caption latency, verbatim captions, word substitutions and paraphrasing, which all may be necessitated when users are unable to read as fast as words are spoken, or when there are multiple overlapping speakers [41]. Ultimately, it depends on these trade-offs as to whether captions are usable by the intended target audiences.

A multitude of quality metrics have been proposed in the past, which are covered in detail in Section 1.2. However, none have gained traction in the United States, primarily due to enduring disagreements across consumers with disabilities, captioning providers, content creators, and broadcasters. Generally, from a user perspective, captions that have an objectively low error rate, may nevertheless not be usable, because they may be too fast. From a caption provider perspective, similarly, captioners may not be able to keep up with too-high rates of speech, or they may get penalized for minor errors that do not impact comprehension [19]. To break the Gordian knot of the deeply entangled disagreements on caption quality metrics, there is a need for evaluating metrics from two disparate standpoints. On the one hand, TV captioning metrics should measure errors objectively, and at the same time distinguish between severe errors that impact understanding and minor errors. On the other hand, metrics also should capture the user experience – if the user experience is poor, or the user is unable to follow a captioned video, it does not matter if a metrics assign high quality scores, and there are objectively few errors.

This paper aims to provide a deeper understanding of captioning quality from a user experience perspective, and its relationship to quality metrics. We specifically focus on live captions produced for TV broadcasts in the United States. They are generated either simultaneously with the broadcast, or with a few hours of lead time. This mode of production is distinct from offline captions, where the captioner can make multiple editing passes, leading to well-considered decisions about the pacing and placement of captions, speaker identification, and elimination of errors. Live captions, on the other hand, due to time constraints, are much more susceptible to usability problems and severe errors.

In this paper, we present two successive studies that had users rate the quality of captioned video clips taken from TV broadcasts, and compared their ratings with a number of proposed caption quality metrics: Word Error Rate (WER), Weighted Word Error Rate (WWER), Automated Caption Evaluation and its successor (ACE/ACE 2); see Section 1.2.3 for details on these metrics. The first study was a pilot replicating past work on WER and ACE [40] using the WWER metric [4] instead. It also used offline-captioned videos as a control against live captions (Section 3.1). This pilot yielded poorer-than-expected user ratings for the offline-captioned videos, raising questions about how well DHH users can evaluate caption quality to begin with. These questions led to the design of a follow-up study to investigate caption usability in depth, and controlling for factors that may have influenced the ratings in the pilot (Section 3.2). This follow-up study also applied all four metrics mentioned above: WER, WWER, ACE and ACE2.

The main research questions addressed in this paper are:
- (RQ1) What is the relationship between the user experience and caption quality metrics; that is, can quality metrics stand in for testing captions with users?
- (RQ2) Are some metrics better suited than others to capture the severity of different types of caption errors, and is this reflected in user ratings?
- (RQ3) What factors affect the user experience of live TV captions?

In the following, we provide an overview of captioning terminology (Section 1.1), the four quality metrics applied in our studies (Section 1.2), discuss other related work (Section 2), describe the two studies and the results (Sections 3.1 and 3.2), and conclude with limitations (Section 4) and overall takeaways (Section 5).

## 1.1 Captioning Terminology

In this section, we define some key captioning concepts that are used in our study design.

*1.1.1 Live vs Offline Captions*

Live captions are created in real-time during a live event, such as a news broadcast. A human captioner transcribes the audio into text, which is shown to viewers in real time. Offline captions are created either before or after an event and then added to the video.

*1.1.2 Caption Styles*

Roll-up captions scroll up on a fixed region of the screen, typically the bottom. The captions display several lines of text, and often individual words appear as the audio is being spoken. In the United States they are used almost exclusively for live captions in TV broadcasts, as it frees captioners from the need to time their appearance. They are often delayed relative to the audio, reflecting the time it takes to process the sound and produce captions.

Pop-on captions are timed to appear and disappear in synchronization with the audio, a few lines at a time. In the United States, these are used for prerecorded TV shows, and are the main captioning style found on streaming video providers.

## 1.2 Caption Metrics

In this section, we describe the captioning metrics that we have evaluated across our studies in this paper.

*1.2.1 Word Error Rate (WER)*

The simple word error rate (WER) metric is defined as (S+D+I)/N, where S is the number of substitutions where one incorrect word is erroneously put in place of the correct word, D is the number of deletions, i.e. erroneous omissions of words that were spoken, I is the number of insertions of spurious words that was not spoken, and N is the total number of words actually spoken. The calculation of WER has been standardized by the National Institute of Standards and Technology and is a popular ASR evaluation metric [40]. It does not consider whether some words may be more important than others to the meaning of the message.

Canada adopted WER as a caption metric, and promptly ran into resistance by caption providers. It also became quickly apparent that the targeted WER metrics were not achievable for live broadcasts. Canada retracted this rule in 2016 [10] and proposed the evaluation of a modified NER model, developed based on work by Romero-Fresco [36], which also has been adopted in the United Kingdom [18]. NER was developed specifically for re-speaking live captions [30][14] (where a captioner listens to the audio and re-voices it to a custom ASR system). More recently, its accuracy has also been evaluated on steno and ASR captions [35]. While some ASR systems were able to match the quality thresholds for human captioners, the authors note that even when they reach these, the "experience they provide for the viewers is significantly worse than that of human captions." Because of the large manual scoring element involved, and the possible need to adapt specifics to the US market similar to what was done for Canada, NER is beyond the scope of this paper and its validation for our stimuli will need to be addressed in future work.

*1.2.2 NCAM's Weighted Word Error Rate (WWER)*

The National Center for Accessible Media (NCAM) at Boston-based TV station WGBH proposed a weighted word error rate (WWER) for capturing the relative importance of types of errors in captioning. It is a modified version of WER, where errors are assigned a severity weighting [4], developed from caption user surveys. These were developed via surveys carried out with caption users, who rated the severity of errors in stimuli.

WWER also was discussed in the FCC's proceeding on caption quality but failed to gain traction with both the consumer and industry sides. One particularly compelling aspect of the WWER metric is that it can be approximated in a quasi-automated manner using Nuance's ASR engine. Unfortunately, a patent dispute led to the ASR method becoming unavailable, and at present, applying this metric requires human raters. The same dispute also led to a loss of the exact weights used for each error type. In the absence of these, we approximate them by using the error severity percentages from the survey results underlying WWER, as per a recommendation from NCAM[1].

*1.2.3 Automated Caption Evaluation (ACE and ACE 2)*

Kafle et al., [21] developed Automated Caption Evaluation (ACE), which calculates two measures: (1) a word importance prediction score using entropy of a word given its context, as a measure to identify key words in a text and, and (2) a semantic distance between the error words and actual words in the reference text, as an approximation of the deviation in meaning due to errors. ACE combines them in an impact score, which is used to predict the usability of the captions. The original ACE work applied an n-gram-based model to calculate the word importance score. This method was refined in a subsequent publication to use neural network-based language models, to leverage more contextual information and to improve the estimate of word importance. It was released as the ACE 2 metric [22].

Unlike WWER, ACE and ACE 2 determine the weights and impacts of errors fully automatically. However, they were specifically created for evaluating the accuracy of ASR errors in meetings and classroom settings. An open question has been whether they are applicable to live TV captions. Furthermore, it has been unclear whether it is neutral with respect to the captioning method, irrespective of whether human captioners were involved. Prior work by Wells and colleagues [40] did not find a clear advantage for using ACE on human-generated live TV captions, albeit for a small sample size. Note that this paper provides further evidence toward answering these questions (cf. Section 3.2).

Another limitation of ACE and ACE 2 is that it has been designed to work on the sentence level. This means that sentences on a ground truth transcript and a captioning transcript must be aligned prior to running ACE calculations. This alignment can be challenging if the captioner omitted or paraphrased entire sentences, or punctuation diverges from the ground truth. At present manual intervention is needed for the alignment. There also is a need for further research on how to calculate an aggregate ACE/ACE 2 score on videos that contain multiple sentences. For this paper, we follow a recommendation by the authors of ACE/ACE 2 to calculate the arithmetic mean across all sentences in a transcript[2].

## 2 RELATED WORK

We present our analysis of prior literature in this section, covering prior work on relevant topics such as latency, condensing speech into captions, the challenge of split attention, and user experience research in this area.

### 2.1 Latency and Caption Quality

Aside from accuracy, latency also has a large impact on perceived caption quality. A study by BBC showed that accuracy and latency are perceived differently depending on whether the viewer can hear the words spoken on TV [5]. If a person utilizes their hearing, their tolerance for errors and high latencies decreases dramatically. Presentation rate also has been shown to have an impact [10][14]. The needs of viewers can be diverse, based on demographics and literacy, which suggests that a one-size-fits-all approach may be difficult to achieve. In addition, predictability of captions is related to context and topics. Current metrics do not apply topic models [11][34], although ACE 2 does consider context, as explained in the previous section.

---

[1] NCAM, personal communication
[2] Kafle and Al-Amin, personal communication

## 2.2 Condensing Captions from Speech

Captioners are encouraged to condense the original transcription to provide time for the caption to be completely read and to be synchronized with the audio [12]. This is needed because, for a non-orthographic language like English, the length of a spoken utterance is not proportional to the length of a spelled word. Furthermore, readers need more time to read and understand the context, as captions condense what was said -- markers like accent, tone, and timbre are stripped out and represented by standardized written words and symbols. Faster dialogue has resulted in lower comprehension compared to slower rates [37][38]. To allow readers to keep up, captioners often omit non-speech information, and are encouraged to use common, high frequency words.

Another factor that necessitates condensing captions is that many DHH viewers are not fluent readers. Professional captioners reduce caption delivery rates by paraphrasing utterances and eliminating redundancies. This can reduce mental effort and increase video programming enjoyment for non-fluent readers. As a result of applying all these condensation methods, the real-time caption speed is typically slower than the equivalent speaking speed [19][20]. Television captioning guidelines were developed for an "average deaf or hard of hearing person" reading at 140 words per minute. Their hearing peers, in contrast, keep up with regular speaking rates of 170 words per minute [26][27][28].

## 2.3 Split Attention Effort

Deaf viewers expend more effort to divide their attention between watching two visual sources of input (the captions and any visual elements on the screen) [31][32]. The extra effort in reading captions often causes fatigue and makes viewers less attentive. They are also likely to miss information that requires both seeing the scene as well as reading captions. For example, some viewers miss visual gags, or fail to identify who is the current speaker on-screen.

Previous research has shown that deaf viewers spend more time reading captions than hard of hearing or hearing viewers and many deaf people have lower English fluency compared with their hearing peers [1][2][3]. They may find captions to be grammatically complex and difficult to understand and experience more frustration with reading and understanding them. In fact, deaf people navigating a site in search of information are more likely to leave the search task incomplete [33]. Moreover, when captions are displayed more quickly than a reader can read them, or when the subject matter is unfamiliar, the viewer spends more time processing the information. When watching TV, deaf viewers' primary concern typically is to be able to enjoy television programs. This means that a full transcript (verbatim captions) of what hearing viewers receive, at the expense of being able to take in the visual action, is not a main priority, regardless of hearing loss, age or literacy level [39].

## 2.4 User Experience and Captioning

Other prior work has focused on the user experience of DHH individuals with captioning technologies through both automatic and human generated captions for speech and non-speech information. One study investigated automatic versus professional subtitles and found that the latter correlated with better understanding. However, automatic captioning had the potential to rival professional captions with small improvements [24]. Kawas et al. found that while students expressed interest in automatic captioning technologies in the classroom, they remained frustrated with errors, preventing their more widespread use [23]. Other work has evaluated methods for augmenting captions. Berke et al.'s study investigated marking-up captions by highlighting or otherwise indicating that the system is not confident in the accuracy of these words, thereby giving more information about possible caption errors [6]. The study found that while displaying more information could be useful, it can also be distracting, which is yet another example of trade-offs that must be made when considering new features for captioning.

For non-speech information, prior work has investigated incorporating information about music and other sound effects, into closed captioning [29]. User responses to such features vary: For example, [29] found that hard of hearing participants were more receptive to such features than deaf participants, who sometimes had negative reactions. Another research study investigating dynamic subtitles, where subtitles change position on the screen also showed disagreement among hard of hearing and deaf users, with deaf participants generally liking the dynamic captions more than the hard of hearing participants [9]. Prior work has also investigated preferences for on-screen captions versus transcripts for access to online videos [25], and found that preferences were context-specific. Users preferred transcripts for more technical (i.e., unfamiliar) vocabulary, as they have more time to read, process and catch up. Another study showed that DHH participants prioritize certain genres of videos that must be accurately captioned (such as news and politics), and poor captions in these more adversely affect their viewing experience [7]. These results show the need for individualized settings for captions – DHH people have varying caption display preferences, and these preferences are often context-specific.

## 3 CAPTION QUALITY STUDIES

In this section, we describe the two studies carried out to address the research questions described in the introduction. The first study (Section 3.1) was a pilot aimed at addressing specifically RQ1 and RQ2: the relationship between the user experience and caption metrics, as well as whether some metrics capture the user ratings better than others. The results of the pilot raised questions about whether participants can tell the difference between live TV captions and offline captions, which led to the addition of RQ3 for the second study (Section 3.2): what are the factors affecting the user experience?

Both studies worked off the same fundamental premise. Study participants were exposed to a counterbalanced succession of short, captioned video clips captured from TV programming and asked to rate the caption quality for each. Each video clip ranged from 23 to 57 seconds. We also calculated the caption quality metrics for the TV captions captured with each of these clips and compared these to the participants' ratings.

### 3.1 Study 1: Pilot with Live TV Captions Using WER and WWER

The purpose of this pilot study was to address RQ1 and RQ2 via replicating the methods of a prior study that assessed ACE [40], in order to assess the utility of WWER.

*3.1.1 Participants*

In total, 22 local DHH participants completed the pilot study. Most were in the 18-34 age range, with the remainder being between 35 and 64 years old. The majority self-identified as Deaf (18 out of 22 participants). Furthermore, there was a nearly equal male-female split. Most self-identified as White, Asian or Other, while two participants self-identified as Black, and one self-identified as Native Hawaiian/Pacific Islander. Over three-quarters of the participants indicated that they use captions multiple times per day or every day, and almost all others indicated that they use captions at least two to three days per week.

*3.1.2 Materials*

As mentioned above, all stimuli were clips captured from locally broadcasted TV programming. The captions were extracted as SRT files from the data embedded in the video captures. Each video was additionally assessed for transmission errors, which are garbled words caused by weak cable signals, and excluded if these occurred. In total, we produced 20 short video clips from four full-length videos: a documentary on the musician Kurt Cobain, a football highlights reel, an episode from ABC News, and a special Town Hall event featured by CNN News. The choice of clips was guided by having a diversity of topics, while avoiding content that may be perceived as overtly political or religious.

For each clip, we produced two separate caption files. The first contained the live TV captions at the same time codes as the video itself; that is, any caption delays present in the TV captions were also present in the clips. The second contained offline captions generated by Rev.com synchronized with the audio. The offline captions were considered the ground truth (i.e., 100% accurate) and used as the reference transcripts for assessing caption accuracy. To obtain our 20 stimuli videos, we extracted 5 clips from each of the 4 full-length programs, and in each set of 5 clips, 4 featured TV captions, and 1 featured captions produced by Rev.com. In total, there were 16 videos with captions recorded from TV broadcasts, and 4 ground truth videos with Rev.com captions to serve as controls. All videos had the audio track removed to remove potential confounders with participants listening to audio.

To control for caption presentation factors, we presented all captions in the roll-up style, even those from Rev.com, since this is how they were originally shown on TV. We used a custom video player to replicate the exact appearance of TV captions in our video stimuli, with no audio and smooth roll-up captions in the default font size stipulated by United States TV captioning standards [12][15], featuring white text on a black background.

*3.1.3 Method*

Participants were given informed consent to review and sign, with the option to have it relayed through sign language. They were compensated $25 in cash or Amazon gift card, and given 60 minutes to conmplete the study. The study was conducted in-person. The participants were seated in front of a 27-inch computer monitor and asked to watch the 20 different videos via a web-based survey in randomized order. Participants were not informed of the existence of controls at all. The stimulus video playback was forced into full-screen mode, with the videos hosted on a local server with a high-speed internet connection to avoid playback artifacts related to bandwidth limitations.

Immediately after each video, the survey prompted them to answer three questions: (1) how the participants rated the quality of the captions, (2) how much of the content did they understand (3) whether they did notice errors in the captions. The first two questions had associated 7-point Likert scales, and the third one had a yes/no response.

To determine the WER and WWER for each stimulus with TV captions, we filtered the TV and ground truth Rev.com SRT files into transcripts without time codes, punctuation and other non-alphanumeric characters. We also removed speaker identification labels and non-speech information (e.g., "[Laughing]"), as these are not considered to be a part of any of the metrics under discussion in this paper. We used NIST's SCLite software[3] to calculate the WER of these transcripts. SCLite also provided detailed word alignments and errors vis-a-vis the ground-truth transcripts. A team, of three raters evaluated each of the aligned errors and assigned them the appropriate WWER classification code following NCAM's work. We then calculated the weighted overall score as described in Section 1.2.2.

*3.1.4 Results*

The participants' ratings of caption quality and understanding are shown in Figure 1 for both the live TV captions and the ground truth captions. Figure 2 shows to what extent participants noticed errors in each video.

---

[3] https://github.com/usnistgov/SCTK

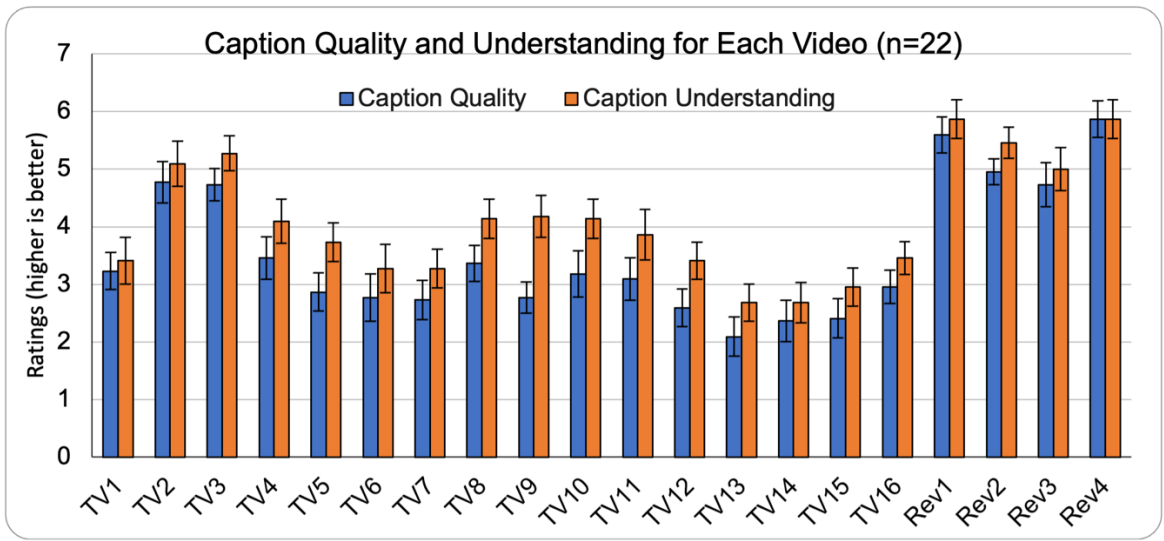

Figure 1: Participants' mean ratings of caption quality and understanding of captions for both videos with TV captions (labeled with the TV prefix) and ground-truth captions (labeled with Rev prefix). The TV captions were shown as they were shown live, including errors and delays. The Rev captions were professionally made with 99% accuracy.

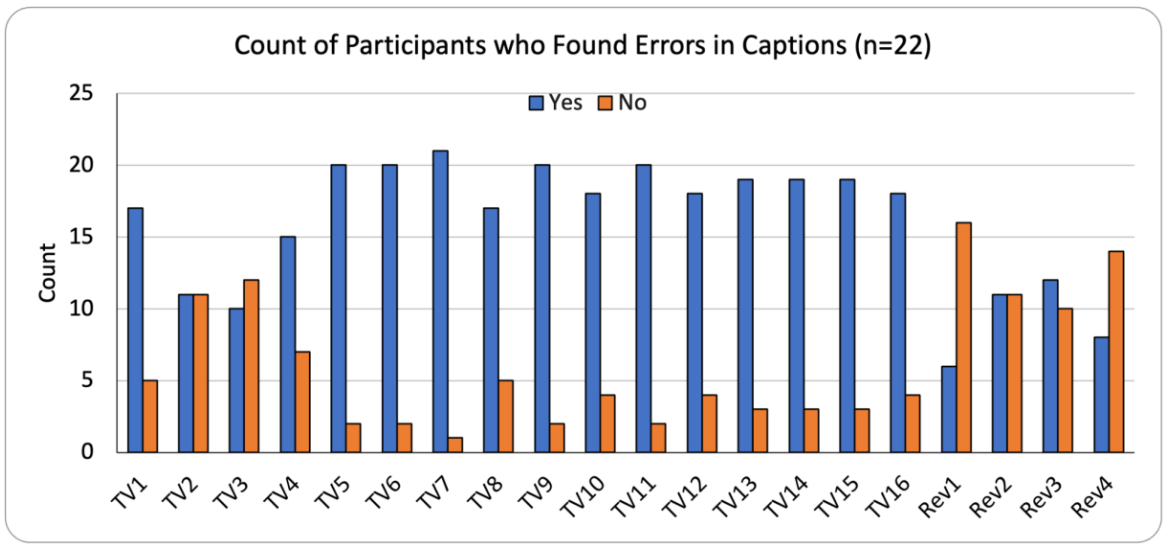

Figure 2: Participants' answers as to whether they noticed errors in the captions for each video. TV captions are prefixed with TV, and ground-truth captions are prefixed with Rev. The TV captions were shown as they were shown live, including errors and delays. The Rev captions were professionally made with 99% accuracy.

The caption quality and understanding ratings for each video greatly varied across participants, with a median standard deviation of over 1.5 points. For the TV captions, the overall average quality score was poor at 3.09 points, while for the ground truth captions, the overall average quality score was 5.28 points, which corresponds to merely "good" on the Likert scale and was only 0.5 points better than the best-rated TV captioning video. In fact, the worst-rated ground truth video was no better than the best-rated TV captioning video. Comprehension averages were higher than quality averages, but almost perfectly correlated with each other (r=0.97). We did not test for statistically significant differences between TV and offline captions, because the pilot was set up to replicate a prior experiment [40]. It was not properly designed for 2-sample statistical comparisons, as the TV and offline clips were distinct from each other, and the difference in their sample sizes was large.

Participants said that they noticed captioning errors in every video, even the ground truth videos. For two of the latter, half of all participants claimed to have noticed errors, which is on par with the best-rated TV captioning videos. For the other two ground-truth videos, only up to a third of participants claimed to have noticed errors.

The median WER score was 24.55% (max 42.86%, min 11.54%), while the median WWER score was 10.56% (max 24.05%, min 1.96%); for both metrics, lower scores mean better quality. WER and WWER scores themselves were tightly correlated (r=0.96). With only TV captions considered, participant quality ratings were weakly correlated with both WER (r=-0.26) and WWER (r=-0.15). Participant understanding ratings had an even weaker correlation (WER r=-0.16, WWER r=-0.06).

*3.1.5 Discussion*

The weak correlations between user ratings and both WER/WWER suggest, for the initial pilot, that these two metrics are unlikely to be a good stand-in for testing captions with users (RQ1). Furthermore, the tight correlation between WER and WWER suggests, as an initial pilot result, that WWER is unlikely to be an appropriate metric for gauging the severity of errors as opposed to simply using WER (RQ2).

Prior to conducting the experiment, we had expected that participants would rate the ground-truth control captions as very good or excellent. The fact that they did not rate them any better than "good" came as a major surprise, which raises important questions about the nature of caption quality assessments. It immediately begs the question as to whether users can discern the difference between live TV and offline captions in the first place. If the answer were that they cannot, trying to correlate quality metrics with the user experience would be futile. To this end it is important to understand better what factors affect the users' experience of live TV captions (RQ3), in addition to testing caption metrics against user ratings. To answer these questions, we conducted an in-depth follow-up experiment, which is described in the next section.

**3.2 Study 2: Offline Captions vs Live TV Captions Using WER, WWER, ACE and ACE 2**

The main purpose of this study was twofold. First, it aimed to address RQ3 directly – what factors affect the user experience with respect to live captions on TV. We employed a mixed methods design to address RQ3 both quantitatively and qualitatively. Quantitatively, a major part of this study was to examine potential confounding factors that may explain the poorer-than-expected user ratings for the ground-truth captions produced by Rev.com in the pilot study. We specifically hypothesized these four possible factors: (1) Users may have expectations of poor caption quality on TV. Roll-up captions are emblematic of live TV captions. Could this caption style have biased participants? (2) English literacy is highly variable among DHH people [1][16][17]. Are user ratings related to literacy? (3) The prior study removed audio. Could the absence of audio have affected user ratings? (4) Many offline caption providers compete on caption quality promises. Could quality differences across providers have affected user ratings?

Second, this study aimed to provide additional data across four caption quality metrics WER, WWER, ACE and ACE 2, in order to strengthen the evidence toward RQ1 and RQ2. The rationale for picking these four metrics was in part that they have been brought up in proceedings in front of the FCC, and in part that they either can be automated or hold the promise of being automated. We excluded NER from the list of assessed metrics while it is still work in progress for the United States captioning market.

*3.2.1 Participants*

We recruited 54 DHH participants, as well as 17 hearing participants as a control group. Self-identification was split across 33 deaf participants, 11 hard of hearing participants, 6 hearing loss participants, and 4 participants who identified as "other." Among the hearing participants, one explicitly self-identified as a child of deaf adults (CODA). The gender distribution across both groups was 21 males, 48 females, 1 non-binary/third gender, and 1 unidentified. The majority of participants identified as White/Caucasian, with 11 identifying as non-White.

The participants' ages skewed toward younger people on one end, and toward older people on the other, with relatively few working-age individuals. Participants were recruited both locally and nationwide across the United States for remote participation. Overall, 21 people participated on-site and 50 participated remotely via Zoom.

Among the DHH group, 23 participants indicated that they often/always watch videos with sound on, and another 11 indicated that they sometimes watch videos with sound on. The remaining 20 participants indicated that they rarely or never watch videos with sound on. We also asked participants to self-rate their proficiency in reading English. Among our DHH participants 43 out of 54 rated their proficiency as "very high", while 10 participants rated it as "high", and only 1 rated as "intermediate". Among our hearing participants, 14 out of 17 individuals rated their proficiency as "very high" and 3 rated it as "high."

*3.2.2 Materials*

As in the pilot study, all stimuli consisted of short video clips that were captured from local TV broadcasts, where the live TV captions were extracted into SRT files and assessed for transmission errors. We selected clips from four different TV shows: The Amazing Race (action and dialogue-focused), CBS News (news and sports), Fox 5 DC (news and sports), and The Real season finale (a talkshow with multiple speakers). Overall, we prepared 9 short self-contained clips per show (about 20-40 seconds long), for a total of 36. To manage participant fatigue, each participant was presented with 24 out of the 36 total clips in a counterbalanced manner.

For each clip, we prepared captions from three different sources. One of these was the broadcast live TV captions, as in the pilot study. The other two sources were captions from two offline captioning providers. These were Rev.com as in the pilot study, and 3PlayMedia, which is a competing entity located in the United States. They both claim a measured accuracy of 99% and employ professional transcribers. We used the Rev.com captions as the ground truth for all caption metrics, like the pilot study. The purpose of adding 3PlayMedia was to control for the possibility that user ratings might have been influenced by a particular choice of caption provider. We note that it is not realistically possible to obtain 100% objective accuracy for captions [41], which is why offline providers only claim 99% rather than 100% accuracy.

Unlike in the pilot study, where we employed only the roll-up captioning style, we used both pop-on and roll-up styles (cf Section 1.1.2). The captions provided by Rev.com and 3PlayMedia had already been timed for a pop-on style by the respective company. We additionally converted them to roll-up style through our custom media player mentioned in Section 3.1.2. For the TV captions, which originated in roll-up style from our broadcast captures, we converted them automatically to pop-on style by combining two successive lines of roll-up text into a single pop-on caption and merging their time codes. The purpose of using both styles was to control for the possibility that users automatically acquire a negative bias when they see the roll-up style.

As in the pilot study, we rendered the captions in the default TV style onto the source videos, with a white font on a black background, for both pop-on and roll-up captions. However, we retained the full audio track in all stimuli. Across all TV captioning stimuli, for both pop-on and roll-up styles, we also adjusted the timing of each caption relative to the audio to eliminate latency as much as possible (in some cases, overlapping speakers required a judgment call). This was done to remove latency as a confounder.

In contrast to the pilot study, we combined the clips (36), caption source (3) and style (2) into a full 3x2 factorial design, where each factor was applied to each of the 36 videos in a counterbalanced manner. There were six conditions in total, resulting in 6x36 = 216 unique video clips. Participants were shown four videos per condition – one clip per TV show –, and across all participants videos were rotated across each condition. Within each condition, the order of videos was further randomized for each participant to avoid ordering effects.

We also used a short English test[4] to score participants' English proficiency beyond their own self-assessment, for which we graciously received permission from the publisher. This test includes two sections on grammar, one section on vocabulary, and one section on reading comprehension. It is designed to be completed in 10-15 minutes and yields an overall numeric score. While this test has not been psychometrically validated, it was the only one that met the criteria of being available online, lasting no longer than 15 minutes, and having permission to use it on participants' computers.

*3.2.3 Method*

*3.2.3.1 Study Procedures*

This study was conducted remotely over Zoom; however, participants local to the area were given the option to participate in the Zoom meeting on-site on a computer with a 27-inch monitor. Participants were paired with an experimenter who was fluent in their preferred mode of communication (American Sign Language, written English via Zoom chat, spoken English via Zoom audio/lip-reading, or a mix). All participant-facing personnel were DHH. Participants were given informed consent to review and sign, and the study took 60 minutes to complete. Compensation of $25 in cash or Amazon gift card was provided.

During the session, participants were asked to share their desktop on Zoom. The screen share allowed us to monitor their progress and to assist in case of technical difficulties. It also ensured that participants received the best possible video quality by playing them back on their computers through their own web browsers, rather than receiving them through Zoom. All sessions were recorded for further analysis; see below.

Prior to the session, participants were informed that they would watch each video only once, and to focus their ratings on the quality of the captions themselves, rather than the fonts, colors, or size. Participants were further instructed that caption quality criteria included good accuracy of captions, speaker identification, no missing captions, timing synchronized with the audio, and that captions do not hide important content.

Like in the pilot study, participants used a web-based survey. They were presented with six sets of four videos each, corresponding to each of the six experimental conditions. As in the pilot study, participants were not given any information on the captioning source or the differences across the conditions. After each individual video, forced to play in full-screen mode by the survey, participants were asked to rate the caption quality on a 7-point Likert scale.

After each set of videos, participants received several follow-up questions asking them about their overall experience with respect to the four videos they had just watched. The first question asked how much of the content they had understood, rated with a 7-point Likert scale (very similar to the question in the pilot study, but asked only once per set of four videos to reduce fatigue). Participants were also presented with an overall matrix question that drilled down into their ratings of the videos. This question had six components, each with a 5-point Likert scale indicating agreement/disagreement with the following statements: (1) The captions had too many errors, (2) The captions were too fast, (3) The captions were delayed, (4) The captions were freezing (5) The captions were difficult to read, and (6) The captions were missing punctuation. Participants answered these Likert questions through an online Qualtrics survey.

---

[4] https://www.transparent.com/learn-english/proficiency-test.html

Participants also received condition-specific open-ended questions. This was done both to collect qualitative data and to break up the monotony of watching and rating videos. Participants could respond in their preferred mode of communication. The questions themselves were drawn from the following: (1) Which, if any, of the four videos you just watched had the worst captions, (2) Did you feel that the captions were blocking any important visual video content in the videos you just watched, (3) Did you feel that you could read the captions comfortably in the videos you just watched, or was there too much text, and (4) Were you able to catch any of the action in the video, or did you focus only on the captions? Questions (3) and (4) were asked twice each to cover both pop-on and roll-up-style conditions, thus bringing the total to 6.

At the end of the session, participants were asked to complete the English test on their web browser, while still sharing their screen. The overall numeric score was recorded as part of the survey. Overall, 66 out of 71 participants completed the test, due to some running out of time.

*3.2.3.2 Qualitative Analysis*

We used the Zoom recordings to perform qualitative analysis on the open-ended questions described in the previous section. Depending on the participant's mode of communication, we used deaf personnel fluent in American Sign Language to create a written transcript of the responses, copied text chat responses, or listened to the audio and cleaned up the Zoom auto-captions. In all cases, the result was a written record of the participants' responses.

Three people, split into two teams, coded these responses using Braun and Clarke's thematic analysis methodology [8]. Braun and Clarke's analysis method is an iterative method for coding data and identifying patterns within qualitative data. This process contains several steps, which we followed: We began by doing an initial pass of our transcripts to familiarize ourselves with the data. Subsequently, each of our two teams did their first round of coding independently, going through the data and systemically generating initial codes from participants' comments. The teams reconvened and collated several candidate themes from the initial codes and did a subsequent coding pass together to agree upon and finalize the codes. They iterated through and refined the list of candidate themes several times to create a final thematic map with supporting examples.

*3.2.3.3 Caption Metrics Generation*

We generated caption metrics scores for each video clip with live TV captions using the Rev.com captions as the ground truth. For WER and WWER, we followed the same process as the one used for the pilot study, with a team of four raters. For ACE and ACE 2, we manually aligned the sentences and calculated the ACE and ACE 2 scores for each clip, as described in Section 1.2.3. As before, we eliminated punctuation, speaker identification and non-speech information before applying each metric. As there were 36 distinct clips content-wise (pop-on and roll-up TV captions are identical for metrics purposes), we also had 36 distinct scores for each metric.

*3.2.4 Results*

*3.2.4.1  Caption quality*

Separating the data into DHH and hearing participants, 2-way repeated measures ANOVA tests were conducted, using caption style (pop-on and roll-up) and caption source ("TV", "Rev", and "3PlayMedia") as the within-subject effects. For each set of 4 videos each, quality responses were averaged to get one value per source per participant. All of the p-values were adjusted using Bonferroni corrections. For the caption quality question ("How would you rate the quality of captions?" Likert-scaled from 1=awful to 7=excellent), shown in Figure 3, the ANOVA showed an significant main effect of source ($F(2,106) = 30.03$, $p<0.001$) for the DHH participants, and no significant main effect of style ($F(1, 53) = 3.43$, $p=0.07$). For the hearing participants, the main effect of source was significant ($F(2, 32) = 29.57$, $p<0.001$), but the main effect of style was not ($F(1, 16) = 0.83$, $p=0.37$). There was no significant interaction effect for both populations (DHH: $F(2, 106) = 2.54$, $p=0.083$; Hearing: $F(2, 32) = 0.78$, $p=0.46$).

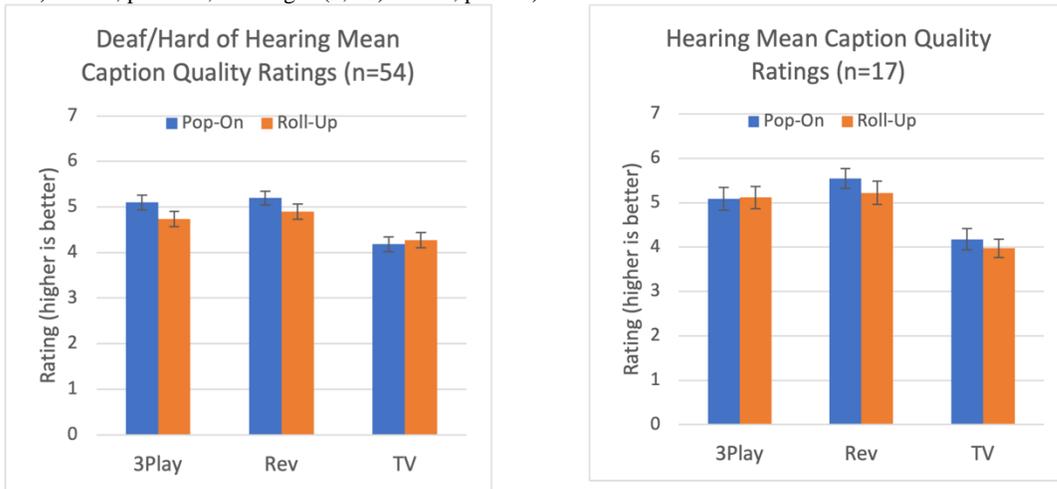

Figure 3: Caption quality ratings for DHH and hearing people. Both groups exhibited a significant main effect for caption source, but no significant main effect for caption style, and no interactions. The TV captions were shown as they were shown live, including errors and delays. The Rev and 3Play captions were professionally made with 99% accuracy.

For DHH participants, post-hoc comparisons using Tukey HSD test show that the mean quality rating for TV captions were significantly lower than both Rev captions ($t(53) = 6.812$; $p<0.0001$) and 3-Play captions ($t(53) = 5.938$; $p<0.0001$). The Rev and 3-Play conditions did not differ significantly ($t(53) = 1.222$; $p=0.4458$). Hearing participants had similar results, post-hoc comparisons showed that the mean quality rating for TV captions were significantly lower than both Rev ($t(16) = 6.408$; $p<0.0001$) and 3-Play ($t(16) = 5.549$; $p=0.0001$). Rev and 3-Play did not differ significantly ($t(16) = 1.962$; $p=0.154$).

Using Pearson's correlation coefficient, the correlation between participants' English test scores and caption quality ratings were calculated. For the DHH subjects, all correlations were weak; $r=0.149$ for TV, $r=0.234$ for Rev, and $r=0.159$ for 3PlayMedia. For the hearing subjects, correlations were similarly weak; $r=0.190$ for TV, $r=-0.239$ for Rev, and $r=-0.246$ for 3PlayMedia.

*3.2.4.2 Comprehension*

Using the same analysis as for caption quality, with scores again averaged over four videos to get one comprehension rating per source per participant, the main effect of source was significant for DHH (F(2, 106) = 7.46, p<0.001) and significant for hearing (F(2, 32) = 6.805, p=0.003), see Figure 4. The main effect of style was not significant for either (DHH: F(1, 53) = 1.46, p=0.23; Hearing: F(1, 16)=0.119, p=0.74). There was also no interaction effect for both (DHH: F(2, 106) = 1.40, p=0.25; Hearing F(2, 32) = 0.246, p=0.78).

For DHH participants, post-hoc comparisons using Tukey HSD test show that the mean quality rating for TV captions were significantly lower than both Rev captions (t(53) = 4.049; p=0.0005) and 3-Play captions (t(53) = 2.418; p=0.0492). The Rev and 3-Play conditions did not differ significantly (t(53) = 1.264; p=0.4218). Hearing participants had similar results, post-hoc comparisons showed that the mean quality rating for TV captions were significantly lower than both Rev (t(16) = 2.669; p=0.0421) and 3-Play (t(16) = 3.105; p=0.0177). Rev and 3-Play did not differ significantly (t(16) = 0.416; p=0.907).

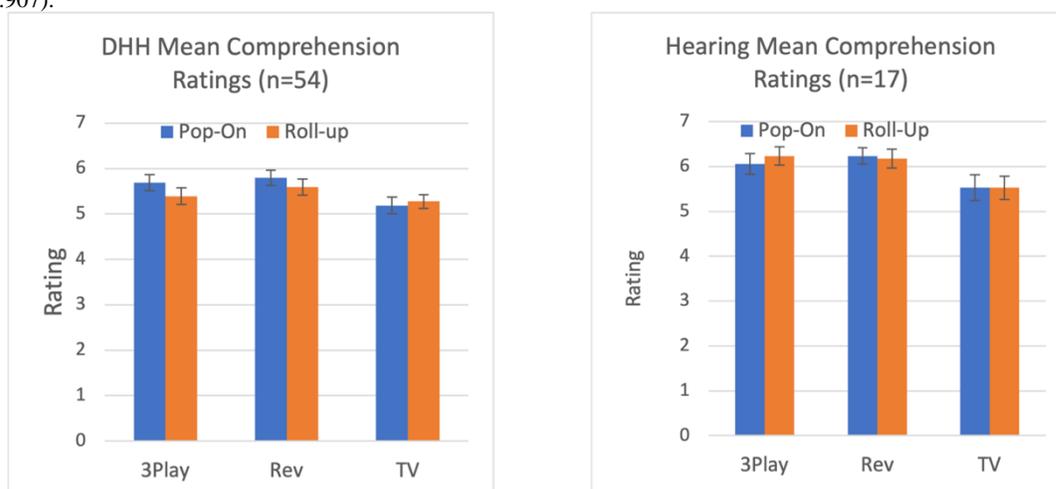

Figure 4: Caption comprehension ratings for DHH and hearing people. Both groups exhibited a significant main effect for caption source, but no significant main effect for caption style, and no interactions. The TV captions were shown as they were shown live, including errors and delays. The Rev and 3Play captions were professionally made with 99% accuracy.

*3.2.4.3 Matrix questions*

Using 2-way repeated measures ANOVA tests (using caption style and caption source as the within-subject effects), for the DHH participants there was a significant main effect of source for the matrix questions about freezing (F(2, 94) = 6.719, p=0.011), many errors (F(2, 102) = 31.18, p<0.0001), hard to read (F(2, 102) = 5.611, p=0.029), and punctuation (F(2, 92) = 11.011, p<0.001). There was no significant main effect of style, except for the hard to read question (F(1, 51) = 9.208, p=0.023). Note that the degrees of freedom differ based on how many participants chose "not applicable" as a response to each matrix question.

For the hearing participants, there was also a significant main effect of source for many errors (F(2,30) = 56.152; p<0.0001), too fast (F(2, 30) = 10.511; p=0.002), punctuation (F(2, 32) = 11.339; p=0.001), and delayed captions (F(2, 32) = 7.885; p=0.010). There were no significant main effects of style for the hearing participants.

We conducted post-hoc t-tests on our results for caption source. For DHH participants, 3Play and Rev captions performed significantly better than TV for the matrix questions about many errors (3Play: t(51) = 6.11, p<0.001; Rev: t(51) = 6.58, p<0.001) and punctuation (3Play: t(46) = 3.82, p=0.0066; Rev: t(46) = 4.31, p=0.0018). Furthermore, Rev performed significantly better than TV for freezing (t(47) = 3.56, p=0.015), but 3Play did not significantly perform better than TV for this question (freezing: t(47) = 2.37, p=0.334). No comparisons between 3Play and Rev were significant.

Post-hoc t-tests for hearing participants showed that 3Play and Rev captions performed better than TV captions for the matrix questions about many errors (3Play: t(15) = 10.04, p<0.001; Rev: t(15) = 6.82, p<0.001). Furthermore 3Play, but not Rev, performed better than TV for too fast (3Play: t(15) = 4.27, p=0.0108; Rev: t(15) = 3.027, p=0.131). Rev, but not 3Play, was significant over TV for punctuation (3Play: t(16) = 3.42, p=0.0552; Rev: t(16) = 3.85, p=0.0228). Neither 3Play nor Rev were significant over TV for delayed captions (Play: t(16) = 3.23, p=0.0816; Rev: t(16) = 3.10, p=0.107). Comparisons between 3Play and Rev showed no significant differences.

Generally, participants indicated that they found all six error types across all conditions (Figure 5), analogous to the pilot study. Using a three-way repeated measures ANOVA, with caption source and style as within-subject factors and DHH and hearing as between-subject factors, only one effect remained significant after Bonferroni correction p-value adjustments for multiple comparisons. For the matrix question about errors ("The captions had too many errors" Likert-scaled 1=strongly disagree, 5=strongly agree), there was an interaction effect between DHH/hearing and caption source (F(2, 132) = 8.69, p<0.001). Hearing subjects on average rated the TV captions as much worse than the 3PlayMedia/Rev captions relative to DHH participants (Figure 6).

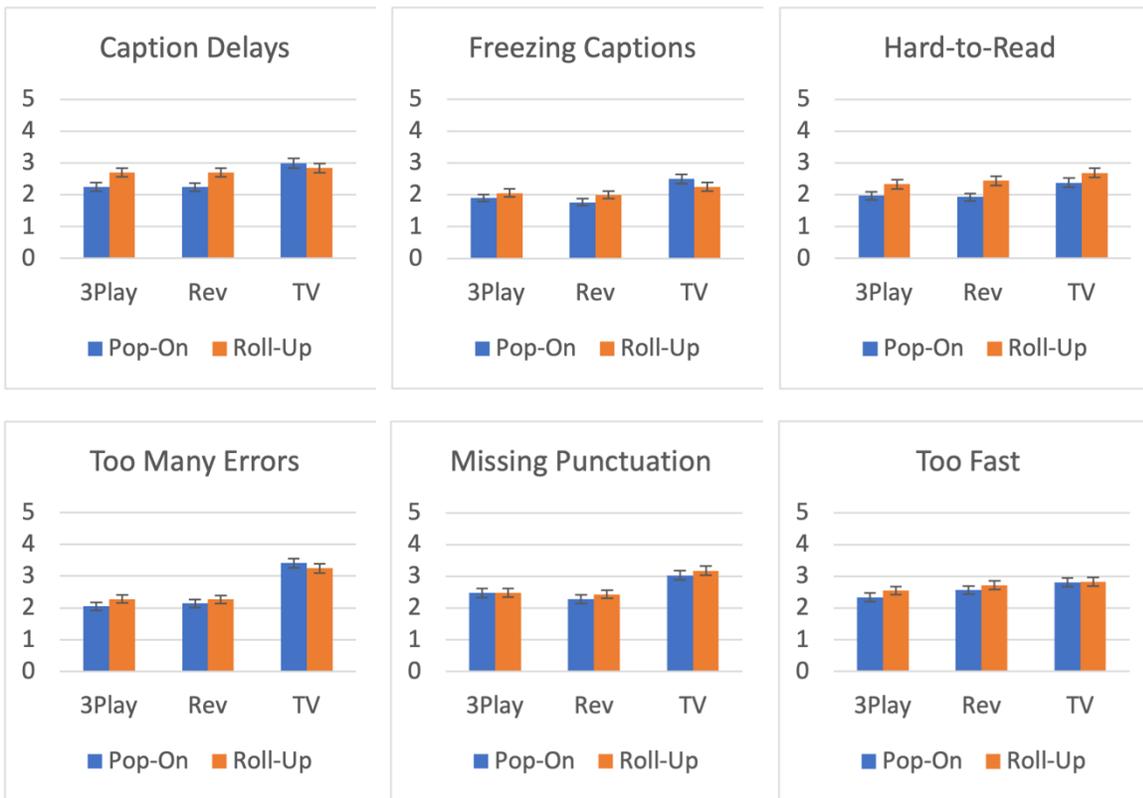

Figure 5: Matrix question results combined for DHH and hearing participants. Participants flagged perceived errors across all conditions and caption sources. Further, there was a significant main effect of source on whether there were too many errors. The TV captions were shown as they were shown live, including errors and delays. The Rev and 3Play captions professionally made with 99% accuracy.

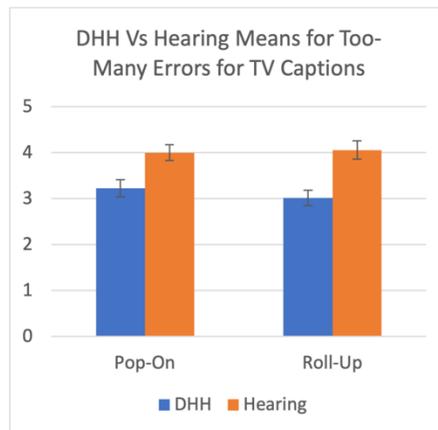

Figure 6: For the question as to whether there were too many caption errors, hearing viewers rated TV captions significantly worse than DHH viewers, exposing an interaction effect in the three-way-ANOVA.

Similar to the caption quality question, Pearson correlation coefficients were calculated for English test scores and comprehension. For the DHH subjects, all correlations were weak; r=0.172 for TV, r=0.238 for Rev, and r=0.282 for 3PlayMedia. For the hearing subjects, again, correlations were similarly weak; r=0.295 for TV, r=0.009 for Rev, and r=0.039 for 3PlayMedia.

*3.2.4.4   Caption Metrics*

Using the WER, WWER, ACE, and ACE2 measurements, we investigated correlations between the metrics themselves, as well as user ratings (see Appendix A for the detailed ratings). Using Pearson correlation coefficient calculations for average user ratings ("How would you rate the quality of captions?" likert-scaled 1=Awful 7=Excellent) shows increasing correlations with WER (r=-0.359), WWER (r=-0.413), ACE (r=-0.412), and ACE2 (r=-0.630) for the pop-on caption style. For the roll-up caption style, ACE2 also had the strongest correlation (WER r=-0.440, WWER r=-0.524, ACE r=-0.438, ACE2 r=-0.528). WER and WWER are very strongly correlated with each other (r=0.956), and similarly for ACE and ACE2 (r=0.917). Neither the user quality rating nor comprehension questions are clearly correlated with the rankings established by the metrics.

*3.2.4.5   Qualitative Analysis Results*

The thematic map of participant responses to the open-ended questions is shown in Figure 7 below. Our analysis indicated that, when assessing caption quality, the three following areas of interest significantly shaped our participants' opinions: (1) Errors within captions, (2) How hard the captions were to follow, and (3) Caption Appearance.

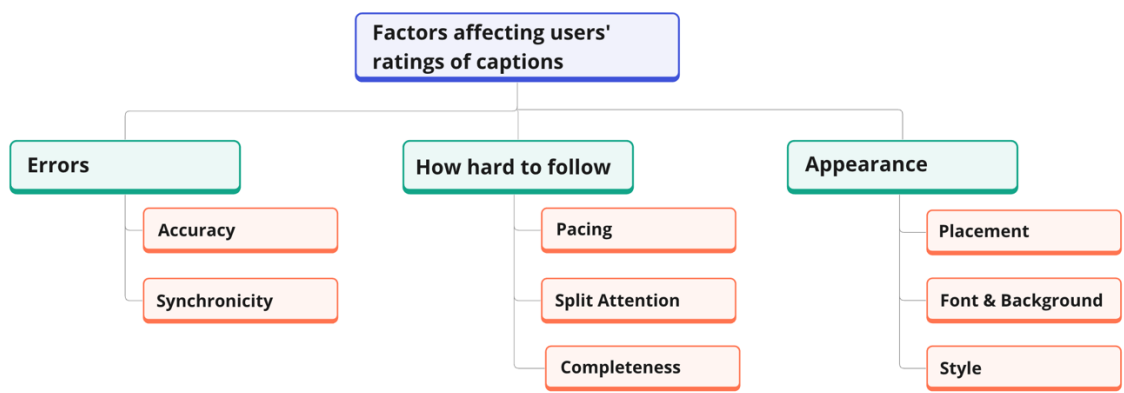

Figure 7: Our final thematic map showing 3 major categories for factors that affected our users' ratings of captions: Caption errors, how difficult they were to follow, and Caption appearance.

**Errors**

Participants often expressed significant frustration when encountering errors in captions. Our analysis revealed that these could be categorized within two subcategories: Accuracy and Synchronicity. Accuracy refers to whether the captions themselves are faithful to the source material and do not have any mistakes or discrepancies. Many participants noted spelling and punctuation errors as well as missing words in the captions. This frustration was especially apparent for those who rely on their hearing or lipreading ability while watching videos: "I don't feel like watching these shows due to caption errors. Many errors showed up on the captions, it did not make sense to me because it did not match the shows" (P 65, deaf) and "The last one [was the worst video] because they were wrong words, and it was missing sentences." (P87, hearing). Several participants commented that they noticed errors in captions, but they thought the captions were of "alright" quality since they have come to already expect poor-quality captions when viewing programs and that it's "impossible" to be perfect, e.g. *"Some had mistakes again, but I don't think it wasn't bad, I could understand what they were saying. I know with live TV, you can't have awesome captions because it is not possible, so….yeah, I thought it was all good."* (P1, deaf).

Synchronicity, or how well the captions line up with the audio, also played a major factor on participants' perceptions of the captions. Generally, participants rated captions more highly when they perceived that captions were in-sync with the audio and with the lip movements of speakers in the video. Interestingly, even though the video stimuli were created to have the captions in sync with the audio, several participants, both deaf and hearing, commented that captions were delayed or ahead: "*Something seemed off in one of the Lugano races, but not sure what. Since I am listening also, it seemed as if captions came on before speech in one"* (P 39, deaf). "*But the timing of the captions in this block was so weird that it was really like distracting in a way that the previous block was not. So it was sometimes it felt like it was on, like the captions were appearing on a timer or something."* (P 77, hearing). While the timing of the audio and captions were synced, these particular sets of videos had been shown with roll-up captions, and it is probable that the exact timing of the word-by-word captions didn't quite exactly match the audio. Thus, even slight differences in audio and caption sync may have a drastic impact on participants' perceptions of caption quality.

### **How hard to follow**

One major emerging pattern was that participants rated caption quality worse when they felt as if the captions were difficult to follow, which subsequently caused them to miss key pieces of information while viewing a program. This feeling of "missing out" can manifest in several different ways: when the captions go by too fast to be able to fully read, when there are many areas of interests occurring on the screen requiring participants to constantly look back and forth, and when captions feel incomplete; for example, participants do not know who is speaking.

Pacing of captions is a factor in how participants rate caption quality. Most participants felt comfortable with the captioning speed and presentation. However, a substantial minority commented that the captions had too much text or were too fast: *"I was focusing only on the captions, it was too fast and it was hard for me to read the lips and follow the captions simultaneously since they weren't synchronized."* (P 26, hearing loss) and *"When it came to the news, it was comfortable for me to read the captions, but when there were activities or events, I was uncomfortable as there was too much text to read."* (P 53, deaf). It is important to note that thoughts on pacing and caption speed may depend on participant demographics; a couple of older participants commented that they read slower and thus have more challenges when captions are fast paced; for example, *"It was fast, but it's because they're speaking fast. And that can be a problem for people to get older, because it takes us a little bit longer to process what's being said."* (P26, deaf).

Participants generally had a tough time when there was a lot going on in the video, with multiple areas of interest that required participants to divide their attention. For example, one participant mentioned focusing on the captions less when there was more action on the screen: *"Well, the first one was FAST. With everything going off – it was so fast seeing the captions, so I gave up watching them and watched the action instead because I couldn't go back and forth."* (P40, deaf).

Deaf, hard of hearing and hearing people alike struggled when captions felt as if they were missing contextual information. This was especially the case for captions without any form of speaker identification, which we had instructed participants to consider as an important quality criterion, even though the metrics assessed in this paper disregard it. Knowing who is speaking is very important for understanding the content when watching a program. Otherwise, it is challenging to follow dialogue. Viewers can become lost, especially in group settings or in fast-paced environments, which can lead to increased cognitive load. *"It didn't identify who the people were, or it was blocked. Who the people were, so *you didn't know who was talking. So that became confusing. If you are only reading the captions, you need to know who was talking."* (P 72, hearing). Some participants mentioned that they used the visual action on screen to identify speakers; something that is possible only if the captions are in perfect sync with the audio: *"I had to figure out who was speaking and look up to the action to see which person was talking. It was hard."* (P 61, deaf). This suggests that the presence of two issues rather than one can have a compounding effect on how participants perceive quality. Lacking speaker identification makes it difficult to know who is speaking, but if the captions are additionally out of sync with the audio, this may drastically worsen their experience.

Another example of missing contextual information occurred when one participant commented that the captions seemed to describe events in the past, and due to possibly not hearing the audio, did not realize that the voice-over itself was referring to a past event: *"The videos were switching to new scenes rapidly while the captions were describing something in the past. I had difficulty connecting them with the video."* (P 33, deaf). This indicates that when captions do not convey everything that is being spoken during a program, this can confuse viewers. The problem here occurred when the captions did not include an explanation that the description was a prior event, even though the audio itself did clarify that. Some participants additionally expressed a desire for more information on non-speech sounds and contextual information: *"And also I would like to see there have noise background like (clapping) or (screaming) on the captions to be able understand the [person's] feeling or behavior."* (P56, deaf).

### **Appearance**

Finally, our analysis revealed that caption appearance (how the captions look, how they are styled, and where they appear on the screen) also factors in how participants perceive caption quality. In our study, we showed participants two distinct styles of captions: pop-on and roll-up. We found that our participants were split on which style is more appropriate. *"The roll up captions were better for action, because I had time to read the captions and catch the actions at the same time. So in some situations, pop on was better, other situations, roll up was better."* (P 40, DHH). *"I felt too much texts, that causes my eyes, like my eyes moves a lot. I don't like that kind of caption; roll-up is not for me."* (P 19, deaf). There was no clear consensus, so customization options might be the best approach.

Furthermore, there were several comments from participants about their preferences for caption size, color, and background. Our analysis showed that participants often had negative perceptions towards captions that they felt were inappropriately sized or colored or had a certain background (opaque vs transparent caption backdrop). Some comments are listed here to highlight the variety in opinion: *"I think the text size was too big hence concealing a lot [of] background video content."* (P20, deaf) in contrast with *"I did not like the small size of captions - longer lines are better but trying to read smaller boxes is very hard."* (P22, DHH), *"I would prefer to have them a bit smaller in yellow font on light blue background."* (P54, deaf) and *"It was nice with the transparent background. It wasn't bold and black."* (P7, deaf). Again, there was no clear consensus, and each participant had their own individual preference.

Finally, caption occlusion, caused by suboptimal placement of captions on the screen, was also a big source of frustration for many viewers. If captions block an area of interest on the screen, this can cause participants to miss out on important things, such as news headlines or the score of a sporting match. *"When I see news, it always blocks. Maybe they could move the [captions] up on the news and add a black screen under them for improving visibility."* (P 21, deaf) and *"All videos had captions blocking visual content, such as news or names of locations. Amazing race they had something on there but captions were covering it so I couldn't read it."* (P42, deaf). In fact, some participants commented that they were frustrated enough to the point that they wished they could turn off the captions to capture some of the visual information on-screen; for example, e.g. *"I missed point one [in the video]; I would turn off the caption to see what, then turn it [back] on."* (P52, deaf).

Overall, the various differing opinions and preferences on caption style and appearance (e.g., roll-up versus pop-on, color, size, background, and placement) highlights the need for a robust customization package with features that allow for changing styles, font, size, location, and so forth. When captioning videos, there is no one-size-fits-all solution, and many participants have strong preferences for how captions should be displayed.

*3.2.5 Discussion*

Our first research question, RQ1, aimed to investigate the relationship between user experience and caption quality metrics, and our second research question, RQ2, investigated if some metrics are better suited than others to capture the severity of different types of caption errors, and whether this is reflected in user ratings. Study 1 suggested that the correlation between subjective user ratings and both WER/WWER was weak, and this in conjunction with a tight correlation between WER and WWER suggests that neither WER nor WWER are likely to accurately account for caption error and both are unlikely to be a good indicator for caption quality. Study 2 in part affirms this finding but showed a moderate correlation between WWER and user ratings. It additionally showed that, similar to WER/WWER, ACE/ACE2 are tightly correlated with each other but still are at best a moderately good indicator for user quality rating or comprehension ratings.

Caption metrics are, in conjunction with policy making, used to set minimum accuracy requirements for broadcast TV. However, since our analysis shows that the correlations between user ratings and caption metric measurements are weak to moderate (RQ1), this means that none of the metrics under consideration in this paper can be used to affirmatively judge which of any two captioned videos will be preferred by viewers (RQ2). This implies that using these caption metrics to guide requirements for broadcast TV could actually be detrimental if they cause TV captions to worsen in quality from the viewer's perspective. Additional work will be needed to develop more accurate metrics that truly reflect the user experience, which in turn will result in more effective TV regulation policies. Our analysis for RQ3 (What factors affects the user experience of live TV captions?) could shed some light on what factors are important and affect users' perceptions of captions and could be useful in testing and developing future metrics, discussed more in-depth in the following paragraphs.

With respect to RQ3, our statistical analysis showed that the caption source has a significant impact on caption quality and comprehension (significant main effects shown in ANOVA tests). The TV captions were rated poorer and yielded lower comprehension. This is true for both DHH and hearing subjects, which shows that the state of current broadcast TV captions need improvement. Statistical analysis also showed that there were no correlations between the English test scores and user ratings for comprehension or quality, which means that user literacy levels are not a primary factor for user judgements. For the matrix question as to whether there were too many caption errors, the hearing participants rated TV captions significantly worse than DHH viewers (the interaction effect was significant in the three-way ANOVA analysis). This implies that people who follow video content via listening notice captioning errors more easily and are more critical of them.

Qualitative analysis revealed that factors that influence our participants' ratings fall under three broad categories: Caption errors, how hard the captions are to follow, and caption appearance. Regarding caption errors, participants generally rated caption quality worse when there were multiple errors (accuracy issues) or when they felt that they were out of sync with spoken information (synchronicity issues). These findings are in-line with prior work, as accuracy and synchronicity are already factors that the FCC considers [15] for industry best practices. These are also supported by our matrix results which showed that users are sensitive to captions being in sync with the audio and video. Even though we controlled latency in this study, current live captioning production methods will always cause a captioning delay irrespective of the methods being used. Future work is needed to assess methods for improving caption accuracy and how to best manage caption latency as even small changes in synchronicity evidently have a big impact on how users perceive caption quality.

Qualitative analysis also showed that, when captions are hard to follow, this significantly impacts viewers' perceptions of quality. A major pain point was the lack of speaker identification. Especially when there was a lot of action on the screen and multiple speakers, it became very difficult for DHH and even hearing users to follow. However, several participants commented that while captions were hard to follow, they recognize limitations and trade-offs between having verbatim captions and caption speed. This finding supports prior work [41] showing that there is no clear solution on how to address these issues as some participants would prefer sacrificing verbatim captions in exchange for a more readable, slower caption speed while others would prefer the opposite approach. This highlights the importance of a robust customization feature, where participants could select precisely how fast the caption would appear or dictate how truncated the captions would be, among other options.

Opinions on caption appearance were overall shaped by caption placement, caption font and background, as well as caption style. Qualitative analysis revealed that a major source of frustration from participants was when captions occluded important visual content, with some desiring to turn off captions completely so they could see what was on the screen. Again, this finding is in-line with FCC guidelines [15], showing it is critical to avoid blocking content, e.g., by moving the captions out of the way when a news headline appears. More work remains to be done on how captions can be best moved out of the way, whether it is via customization, control by the content creator, or automation.

Even though the Rev.com and 3PlayMedia captions were created professionally and are claimed to be over 99% accurate, none of the user ratings (DHH or hearing) for the matrix questions (asking about six caption error types) were significantly below 2 (1=strongly disagree, 5=strongly agree). As we controlled several potentially confounding factors in the experimental design, this result raises important questions regarding the extent to which viewers can recognize whether caption quality is good or bad. They are clearly able to recognize some differences, since TV captions scored worse than offline captions. Yet, the question remains: what would it take for a viewer to give a captioned clip a perfect score? It may be the case that captioning preferences are far too individual to arrive at this point with a one-size-fits all approach. It may be necessary to offer customization options far beyond the font and styling choices available today, to include caption style settings for pop-on vs roll-up, as well as custom-tailoring the content, placement, and speed of captions to the individual.

## 4 LIMITATIONS AND FUTURE WORK

Our participant age ranges were slightly more skewed towards younger ages. Different age groups may have different judgements on caption factors such as being hard to read or too fast, which could be explored in future work. Additionally, all our participants self-rated their English proficiency as high or very high (only one DHH individual chose intermediate). Our analysis showed no relationship between English proficiency and caption ratings, but this study had a majority of people with high proficiency. We do not know if the same would be true for people with low English proficiency, and this can be explored in future work. Additionally, this study may benefit from participants with more diverse backgrounds. Despite targeted outreach, our participants mostly identified as White/Caucasian, with only 19 self-identifying otherwise. It will take an integrated community recruiting effort to capture more diversity.

As mentioned in the previous section, it is imperative to understand what it would take for a user to give captions a perfect rating. Content-based customization will be an important avenue to explore, and future work should allow participants to fully customize their captioning settings. Note that special care must be taken to render roll-up captions correctly, as previously mentioned in Section 3.1.2.

In addition, our short clip duration leaves little time to for the viewer to understand the general topic of the video and get contextual information which can thereby be used to mentally correct some errors in captions.

The study also has not explored user ratings of ASR-generated captions. ASR is becoming increasingly prevalent in many live TV programs across the United States. However, none of the chosen shows for the stimuli featured ASR-based captions, to the best of our knowledge. Future work should explicitly have users rate ASR captions.

## 5 CONCLUSION

Starting with a pilot study, and following up with a fully factorial study, it has been revealed that even high-quality captions are perceived to have errors and problems by both DHH and hearing viewers, despite controlling for potentially confounding factors. Furthermore, our paper analyzed the correlation between user ratings and the assessed quality metrics (which are used in conjunction with policy making to establish minimum accuracy requirements for broadcast TV) and found that the correlation is weak to moderate. This reveals that factors other than caption accuracy may affect subjective user ratings and that using the investigated four quality metrics alone cannot be used as a stand-in for assessing caption quality. This finding has major implications for any definition of "good-quality" captions and raises questions as to what extent metrics can be squared with the user experience of captions. Our follow-up qualitative analysis of open-ended feedback provides some insight as to which factors may be important to users, including but not limited to errors in captions, how difficult the captions are to follow in the context of the video, and the visual appearance of captions on the screen. Furthermore, it appears that caption quality ratings will be inextricably linked to individualized customization options that cover both appearance and content.

**Acknowledgments**

The contents of this paper were developed under a grant from the National Institute on Disability, Independent Living, and Rehabilitation Research (NIDILRR grant number 90DPCP0002). NIDILRR is a Center within the Administration for Community Living (ACL), Department of Health and Human Services (HHS). The contents of this site do not necessarily represent the policy of NIDILRR, ACL, HHS, and you should not assume endorsement by the Federal Government. Additional funding was provided by a National Science Foundation REU Site Grant (#2150429). Norman Williams supported the technical setup of the experiments, wrote the custom video player used for the stimuli, and organized the collection of TV recordings. James Waller provided extensive consultation on the statistical analysis.

## A APPENDIX

The following table shows the metrics for all videos from the Study 2 across WER, WWER, ACE and ACE2. The median WER was 22.45; the median WWER was 8.00; the median ACE was 0.35; and the median ACE 2 was 0.39.

Table 1: Metrics for all videos juxtaposed with mean user caption quality ratings for pop-on and roll-up styles. For all metrics, lower means better. WER ranges between 0-100 (although pathological cases could result in values >100); WWER has no defined upper bound; ACE and ACE 2 are values between 0 and 1. For the mean quality ratings, higher means better on a scale of 1-7.

| Sequence # | Source | WER | WWER | ACE | ACE2 | Pop-On Rating | Roll-Up Rating |
|---|---|---|---|---|---|---|---|
| 1 | Amazing Race | 9.1 | 4.48 | 0.18 | 0.23 | 5.67 | 5.44 |
| 2 | Amazing Race | 27.5 | 6.97 | 0.54 | 0.50 | 5.33 | 4.56 |
| 3 | Amazing Race | 13.9 | 4.67 | 0.16 | 0.19 | 5.50 | 4.29 |
| 4 | Amazing Race | 20.3 | 6.74 | 0.31 | 0.39 | 4.60 | 5.13 |
| 5 | Amazing Race | 19.6 | 6.90 | 0.28 | 0.34 | 5.43 | 4.43 |
| 6 | Amazing Race | 16.4 | 8.59 | 0.24 | 0.37 | 4.67 | 4.11 |
| 7 | Amazing Race | 23.4 | 7.99 | 0.35 | 0.38 | 4.22 | 4.50 |
| 8 | Amazing Race | 23.5 | 10.46 | 0.31 | 0.38 | 4.00 | 2.00 |
| 9 | Amazing Race | 21.7 | 10.53 | 0.35 | 0.31 | 4.33 | 3.44 |
| 10 | CBS | 12.8 | 5.13 | 0.21 | 0.32 | 4.00 | 5.00 |
| 11 | CBS | 17.1 | 4.46 | 0.22 | 0.43 | 3.00 | 3.75 |
| 12 | CBS | 10.8 | 3.30 | 0.23 | 0.33 | 3.00 | 5.57 |
| 13 | CBS | 13 | 3.22 | 0.24 | 0.33 | 4.50 | 4.22 |
| 14 | CBS | 12.7 | 2.97 | 0.25 | 0.34 | 5.63 | 5.57 |
| 15 | CBS | 9.6 | 2.52 | 0.17 | 0.29 | 5.11 | 5.56 |
| 16 | CBS | 8.5 | 3.47 | 0.19 | 0.29 | 5.20 | 4.50 |
| 17 | CBS | 7.4 | 3.99 | 0.23 | 0.21 | 4.00 | 4.00 |
| 18 | CBS | 15.5 | 6.77 | 0.23 | 0.29 | 3.78 | 3.22 |
| 19 | Fox News | 42.6 | 17.47 | 0.71 | 0.76 | 3.33 | 2.67 |
| 20 | Fox News | 37.4 | 17.46 | 0.54 | 0.68 | 2.22 | 3.75 |
| 21 | Fox News | 28.4 | 10.42 | 0.54 | 0.74 | 2.17 | 3.44 |
| 22 | Fox News | 51.9 | 20.50 | 0.76 | 0.76 | 5.38 | 4.57 |
| 23 | Fox News | 52.6 | 25.84 | 0.77 | 0.97 | 1.17 | 2.44 |
| 24 | Fox News | 22.2 | 7.15 | 0.45 | 0.68 | 2.43 | 1.20 |
| 25 | Fox News | 32.4 | 11.91 | 0.52 | 0.56 | 3.57 | 4.43 |
| 26 | Fox News | 43.5 | 18.26 | 0.56 | 0.78 | 3.00 | 3.88 |
| 27 | Fox News | 22.7 | 8.01 | 0.45 | 0.55 | 3.44 | 3.56 |
| 28 | The Real | 45.1 | 15.32 | 0.65 | 0.58 | 5.00 | 4.22 |
| 29 | The Real | 30.9 | 11.65 | 0.59 | 0.60 | 4.00 | 5.00 |
| 30 | The Real | 32.5 | 12.54 | 0.57 | 0.58 | 3.83 | 4.11 |

| Sequence # | Source | WER | WWER | ACE | ACE2 | Pop-On Rating | Roll-Up Rating |
|---|---|---|---|---|---|---|---|
| 31 | The Real | 17.2 | 4.49 | 0.34 | 0.40 | 5.17 | 5.67 |
| 32 | The Real | 45.7 | 22.54 | 0.67 | 0.77 | 4.00 | 1.60 |
| 33 | The Real | 27.7 | 12.11 | 0.45 | 0.47 | 3.14 | 4.29 |
| 34 | The Real | 34.7 | 16.35 | 0.46 | 0.37 | 5.50 | 3.71 |
| 35 | The Real | 29.2 | 10.39 | 0.48 | 0.56 | 4.20 | 4.63 |
| 36 | The Real | 16.9 | 5.09 | 0.17 | 0.20 | 5.44 | 6.00 |